\documentclass[twocolumn, showpacs, preprintnumbers, amsmath, amssymb,
superscriptaddress]{revtex4}
\usepackage{graphicx}% Include figure files
\usepackage{bm}% bold math1

\begin{document}

\title{Intermittency as a universal characteristic of the complete chromosome
DNA sequences of eukaryotes: From protozoa to human genomes}

\author{S.~Rybalko}
\email{rybalko@polly.phys.msu.ru}
\thanks{corresponding author}
\affiliation{Universite de Franche-Comte UMR CNRS 6174, route de Gray 25030
Besancon, France}

\author{S.~Larionov}
%\email{serglarionov@yandex.ru}
\affiliation{Physics Faculty, Moscow State University, 119992 Moscow, Russia}

\author{M.~Poptsova}
%\email{maria.poptsova@gmail.com}
\affiliation{Institute for Computational Biomedicine, Weill Cornell Medical
College, NY, 10021, USA}

\author{A.~Loskutov}
%\email{loskutov@chaos.phys.msu.ru}
\affiliation{Physics Faculty, Moscow State University, 119992 Moscow, Russia}

\date{\today}

\begin{abstract}
Large-scale dynamical properties of complete chromosome DNA sequences of eukaryotes are considered. By the proposed deterministic models with intermittency and symbolic dynamics we describe a wide spectrum of large-scale patterns inherent in these sequences, such as segmental duplications, tandem repeats, and other complex sequence structures. It is shown that the recently discovered gene number balance on the strands is not of random nature, and a complete chromosome DNA sequence exhibits the properties of deterministic chaos.

\end{abstract}


\maketitle

The last decade witnessed outstanding discoveries of the structure of genomic sequences. New types of large-scale polymorphisms such as copy number variants (CNV), inversions, segmental duplications and gigantic palindromes \cite{Bailey2002} were found and described as fundamental features of DNA sequences. For human individual genomes these polymorphisms
comprise a significant part of the complete DNA sequence, up to 12\% for CNVs
\cite{Consort} and more than 5\% for recent segmental duplications
\cite{Bailey2002}. For each chromosome of human genome such a variation
comprises up to 30\% of the sequence \cite{database}. The noted sequences have
a crucial effect on the dynamics of the chromosome sequence evolution.

At the same time, the nature of fundamental mechanisms of genome functioning,
such as recombination and replication, results from a close interaction with
the processes of regulation of gene expression and superspiralization
\cite{Payen2008}. Acting on the level of the entire genome these processes
should have a large-scale effect on the DNA sequence composition and lead to
$GC$ and $AT$ local asymmetry, but the contribution of each of those remains
unknown \cite{McVicker}. On the other hand, global compositional symmetry, as
it is stated in 2nd Chargaff's parity rule \cite{Chargaff1951}, has been
demonstrated for chromosomes from bacteria to human.

It is well known that duplications of different type are a common property of
DNA sequence evolution \cite {Ohno}. Copy number variants, segmental
duplications and gigantic palindromes compose a class of large-scale DNA duplications in eukaryotic chromosomes. How these polymorphisms coexist with the processes that maintain the regularity of the most important genomic functions and with the global compositional symmetry of chromosomal DNA sequence, and how they are reflected in the complexity of sequence structure poses a question about random and deterministic properties of the complete chromosome sequences.

In our recent paper \cite{Poptsova2009} we examined, by the 2D DNA walk method, large-scale properties of genome sequences of 671 chromosomes of bacteria, archaea, fungi and human. We found that via in silico gene sorting by
strand position one can obtain a completely symmetrical form of 2D DNA walk of
the coding sequences. It was shown that the number of genes on different
strands as well as their total length and the cumulative $GC$-skew are
approximately equal to each other for the most eukaryotic chromosomes. As a
result, the second Chargaff's parity rule, is just as valid for coding
sequences and is totally defined by the gene strand positions. With a
gene-vector model, as it has been obtained in \cite{Poptsova2009}, the majority
of the genes incline to accumulate guanine $(G)$ over cytosine $(C)$ and
adenine $(A)$ over thymine $(T)$.

These results supplement the question about stochastic or deterministic nature
of evolutionary processes that shape up the observable properties of chromosome
sequences, especially, the gene number balance on DNA strands for complete
chromosomes. How can this balance exist without any restrictions in the presence of recombinations and random segmental duplications?

In this Letter we consider the the complete eukaryotic chromosome sequences and
suggest simple deterministic model which describes a wide spectrum of large-scale
properties of real chromosome sequences across different taxa. These include gigantic pseudo-palindromes, several tandem repeats against a background of stochastic-like areas, and other large-scale patterns.The obtained results
explicitly show that DNA sequences should not be random but possess a
property of deterministic chaos.

To analyze the generated chromosome sequences we apply the 2D DNA walk
technique developed in \cite{Larionov2008}. The selection of $A-T$ and $G-C$
co-ordinate axes, respectively, is dictated by the complementarity of chains
and the hydrogen bond balance \cite{Gates1985, Mizni}. Using ideas of the
symbolic dynamics and applying them to a one-dimensional map of an interval,
one can generate symbolic sequences according to a vocabulary of partition.

It should be noted that previously a certain deterministic model was already proposed in  \cite{Nicolay2004}. The observed non-linear oscillations of
the $GC$ content and of the $AT$ and $GC$ skews were compared with the
characteristic features of chaotic strange attractors. For the full eukaryotic chromosome DNA-sequences the correlations of  $GC$-skews only with replicons and with the gene strand position (see \cite{Huvet2007}) are as yet not confirmed and show a more complex dependencet \cite{Larionov2008}. In particular, it is characteristic for the human genome
to have long non-coding $T$-rich subsequences and small islands of $A$-rich and
$G$-rich coding sequences. Human genes have frequent strand permutations and
form islands with high level of occupation density. These blocks mainly consist of single genes or bidirectional transcription pairs \cite{Poptsova2009}, and composition of DNA sequence in these regions has oscillations of high frequency while the major part of a genome is non-coding with gigantic (up to several megabases) low frequency intervals \cite{Larionov2008}.
All these properties imply more complex dynamics than it was discussed in
\cite{Huvet2007}.

Consider a map of the interval $I=[0,1]$ onto itself:
\begin{equation}
  x_{n+1}=f(x_n,a)=x_n+x_n^a \ (\textrm{mod 1)},\ a>1.
 \label{eq0}
\end{equation}
This map is known as a Manneville map \cite{Manneville1980} which has been
introduced to explain the intermittency phenomenon in dissipative dynamical
systems. As it is known (see, e.g. \cite{Gaspard1988}), the map (\ref{eq0})
exhibits three dynamical regimes depending on the parameter $a$: for $1 \leq a
< 3/2$ dynamics is normal in the sense that the fluctuations of the observable
generated by the map trajectory are Gaussian; for $3/2 < a < 2$ dynamics is
transiently anomalous; and for $a > 2$ the dynamics is (L\'{e}vy) anomalous
with the L\'{e}vy index $\alpha =1/(a-1)$.

The Manneville map has been used to describe DNA sequences as a composition of deterministic regimes with long-range correlations and chaotic processes. However, the developed approach was
limited to small DNA sequences, corresponding, in biological terms, to introns
and exons \cite{Allegrini1995}.

To develop the symbolic presentation for a one-dimensional map, let us
subdivide the interval $[0,1]$ into subintervals $[0; 1/8; 1/4; 3/4; 1]$. Then,
the symbolic partition $AGTC$ corresponds to these subintervals. In this case
the sequences generated by the map (\ref{eq0}) will have long regular parts and
chaotic domains, similar to some small DNA sequences in 2D DNA walk. However it
was shown \cite{Larionov2006}, that this model demonstrates a very narrow
diversity of generated sequences.

\begin{figure}[h!]
\includegraphics[scale=0.35]{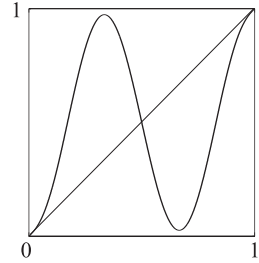}
\caption{Plot of the map with intermittency (\ref{eq1}) with two marginal
points.} \label{Fig1_map}
\end{figure}

To model the real total sequence of nucleotides in chromosomes, let us consider
a modified Manneville map with two accumulating marginal points (traps) and a
complicated partition of the interval based on the triplets of nucleotides
(codons):
\begin{equation}
\label{eq1}
 \begin{array}{ll}
  x_{n+1} = x_n+bx_n^{a}\left(\frac{1}{2}-x_n\right)(1-x_n)^a \
  (\textrm{mod 1)}.
 \end{array}
\end{equation}
In the vicinity of the points $x=0$ and $x=1$ the map has properties of the
original Manneville map (see Fig.\ref{Fig1_map}). Therefore, at certain
parameter values it exhibits the intermittency property with two 'laminar'
regions around points 0 and 1.

\begin{figure}[h!]
\includegraphics[width=1.0 \linewidth]{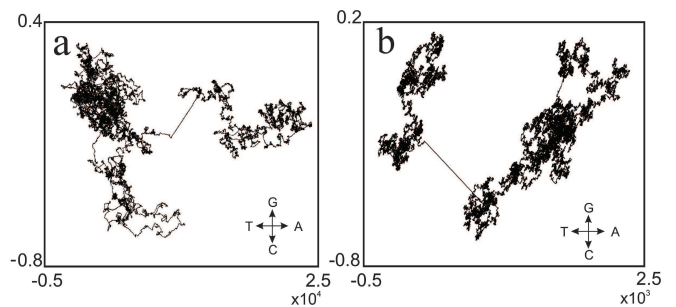}
\caption{(a) 2D DNA walk of sequence of chromosome 4 of Drosophila; (b) 2D walk
generated by the map~(\ref{eq1}).} \label{Fig2_map+Dros}
\end{figure}

Symbolic sequence constructed from $A$, $G$, $T$ and $C$ may be obtained as
follows. Let us divide the interval $[0; 1]$ into 64 equal parts. Each domain
of the partition uniquely corresponds to a single triplet $(s_1, s_2, s_3)$,
where $s_i \in {A, G, T, C}$. Note that the behavior of the phase trajectory
and, as a consequence, the trajectory of 2D DNA walk in this model are
controlled by two numerical parameters and the order of the codons forms a
partition of the interval $[0, 1]$. By varying these parameters one can
significantly change the behavior of the walk. This property of the system
allows us to obtain various types of 2D DNA walk trajectories, qualitatively
reflecting the large-scale characteristics of the real chromosomes.

At the first step let us show that this map allows us to model subsequences
with repeats. The comparison of repeat tracks for the real chromosome and the
chromosome simulated by the expression (\ref{eq1}) are shown in
Fig.\ref{Fig2_map+Dros} a,b, respectively. In terms of symbolic dynamics, for
repeat simulation the phase trajectory should go deep into the `trap' and stay
there long enough to generate corresponding symbols (codons). Long stretches of
tri-nucleotide repeats are a common phenomenon in the real chromosome DNA
sequences. Their origin may be of different nature, but the most common
explanation is that of a slippage during the replication process due to the
formation of stem-loop secondary structures \cite{Mirkin2007}. In our model
such a process is defined by the existence of a long-distance pattern in a
sequence, which represents a path of the phase trajectory deep into the trap. In this model, repeat patterns can also be reproduced in the regime of 3N-nucleototide repeats.

To provide for the `trap' regime, it is sufficient to increase the exponent $a$
in the map (\ref{eq1}). Then if, in the process of iteration, the phase
trajectory comes close enough to the traps (point 0 or 1), it will stay in
their neighborhood for a long time by producing thereby the same codon.

\begin{figure}[h!]
  \includegraphics[scale=0.3]{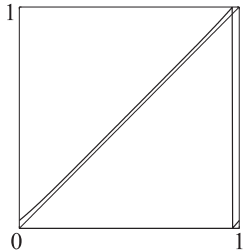}
   \caption{Plot of the map (\ref{ItypeIntre}) with type-I intermittency.}
  \label{Fig3_map}
\end{figure}
\begin{figure}[h!]
  \includegraphics[width=1.0\linewidth]{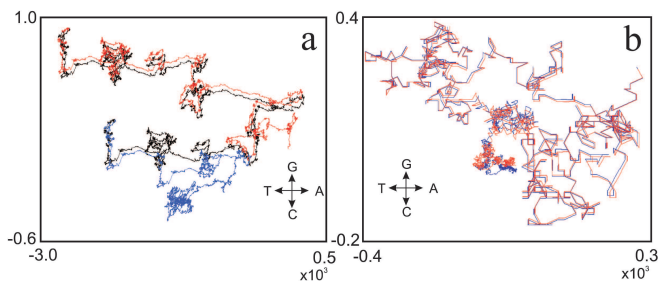}
   \caption{(Color online)
   (a) 2D DNA walk of sequence of duplicated palindrome in human chromosome X
   ($\sim$300Kbs);
   (b) 2D walk generated by the map (\ref{ItypeIntre}). Symmetrical parts of the
   palindrome are marked by red and blue colors.}
  \label{Fig4_map+humpali}
\end{figure}

If one needs to obtain more complex repetitive patterns, one may divide the
interval $[0, 1]$ not into 64, but into a several-fold higher number of
subintervals, e.g. $2\cdot64\cdot n$, where $n$ is an integer. To produce the
inverted patterns it is enough to introduce a symmetry to the specified
partition in such a way that the inverted complementary codons are positioned
symmetrically with respect to the center of the interval $[0, 1]$. This idea
stems from the fact that a real chromosome has two complementary DNA strands
and, as we have shown in \cite{Poptsova2009}, the gene numbers on two strands
are equal for the majority of chromosomes.

In this case we relate DNA strands to 'traps' with the most widely used codons
being positioned in the center of such 'traps'. In terms of this model,
gigantic intervals (pseudo-palindromes) with low frequency changes in
composition can also be interpreted as a result of pseudogenes sequence drift,
which can contribute to the formation of non-coding regions. Importantly, here
the trap, which is close to 0, is out of equilibrium, while the trap close to 1
is a real trap. This property can provide different codon usage statistics that
can qualitatively reproduce a strand-specific mutation pressure in real full
chromosome sequences \cite{Sueoka1992}.

Then in the process of the escaping from traps $0$ and $1$, the phase trajectory
gives rise to statistics of inverse-complementary short patterns, but not the real long inverse-complementary sequences, such as palindromes.

In order to obtain a palindromic sequence for the phase trajectory of the map
it is necessary to go through complementary codons in the opposite directions
immediately after the direct pass of either spacer segments. This can be
achieved when the trajectory is not only out of the traps, as discussed above,
but also enters into them at the same rate as the rate of escape. This means
that the trap should not be a fixed point. Such a configuration can be obtained
with the model which shows type-I intermittency \cite{Schuster1988} that arises
at the destruction of the fixed point through a tangent bifurcation.
\begin{equation}
 \label{ItypeIntre}
  \begin{array}{llr}
    x_{n+1} = & x_n + \left(x_n-\frac{1}{2}\right)^{c}+\epsilon &
   \end{array} (\textrm{mod 1}).
\end{equation}
In a tangent bifurcation there is a narrow region (tunnel) wherein the
trajectory slowly enters and also slowly escapes (Fig.\ref{Fig3_map}). If codons
forming the symbolic partition are positioned complementarily with respect to a
hypothetical center of this area, then, by passing through the tunnel, the
trajectory will create palindromic sequences.

A real duplicated palindrome about 300 kb from human chromosome X is shown in Fig.\ref{Fig4_map+humpali}a. By adjusting the width and profile of the
tunnel in the map (\ref{ItypeIntre}) it is possible to attain a sufficiently
wide variety of generated walks with the inverse-complementary type of
sequences (see Fig.\ref{Fig4_map+humpali}b).

\begin{figure}[h!]
  \includegraphics[width=1.0 \linewidth]{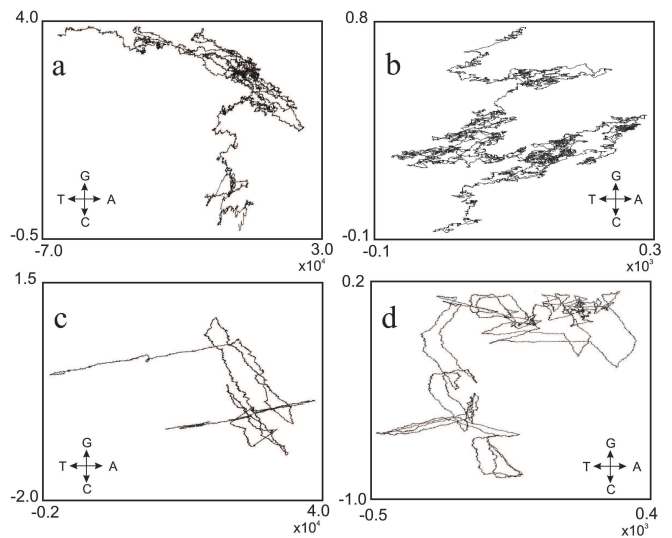}
  \caption{(a) 2D DNA walk of large part of human chromosome X
  ($\sim$30Mbs);
  (b) 2D walk  generated by system of two coupled maps (\ref{2Dcoupled}), $d>0$;
  (c) 2D DNA walk of chromosome 9 of Trypanosoma brucei;
  (d) 2D walk generated by the system (\ref{2Dcoupled}) without
  coupling ($d=0$).}
 \label{Fig5_2Dmap_hum_tri}
\end{figure}

As it was shown \cite{Poptsova2009}, the 2D DNA walk of real chromosomes can be
transformed into a symmetrical trajectory by sorting the genes on strands, and
thus, full chromosome sequences contain a hidden symmetry. Based on the
discovered symmetrical properties of chromosome sequences, we suggest that in
the proposed model the 2D DNA walk of chromosomes can be generally divided into
two types of dynamic processes.

One of the processes is associated with the presence of regions  noted for
their large-scale inverse-complementary symmetry. Such regions may arise as a
consequence of various mechanisms, including inverse intrachromosomal
duplication \cite{Warburton2004, Fairhead2006}, subtelomeres and
telomeres repeats, transposon insertions, recombination-caused strand
switches of genes with close compositional structure, strand-specific
mutational pressure, etc. The second process reflects the fact that in
full-length chromosomes there is a balance in the number of genes on the
strands \cite{Poptsova2009}, and thus this process represents a statistically
confirmed hidden symmetry. As discussed above, the first of the processes can
be effectively modeled by the map with type-I intermittency, which is
characterized by the presence of a narrow `tunnel' that produces the global
compositional symmetry (Fig.\ref{Fig3_map}). The second process is associated
with gene strand switches and can be qualitatively described by the symmetric
Pomeau-Manneville map with two `traps' as shown in Fig.\ref{Fig1_map}.

It is possible to combine these two processes into one two-dimensional model.
The dynamics in one direction is mainly determined by the map with a `tunnel'
and the behavior in the second direction is governed by the map with `traps'.
In general, these two dynamics are mutually dependent. This means that these
maps should be coupled with a system with parametric control of interdependence
of two dimensions. The symbolic dynamics for the constructed two-dimensional
map can be formed by a partition of the two-dimensional space, which is
composed of the direct product of partitions of one-dimensional spaces. The
system of two maps, coupled by discrete diffusion, can be written as follows:
\begin{equation}
  \begin{array}{ll}
   x_{n+1} = x_n+bx_n^{a}\left(\frac{1}{2}-x_n\right)
   (1-x_n)^a +d(y_n-x_n),\\ [4mm]
   y_{n+1} =  y_n+y_n^{c}+\epsilon+d(x_n-y_n),
  \end{array}
 \label{2Dcoupled}
\end{equation}
where $x$ and $y$ are defined modulo 1. Fig.\ref{Fig5_2Dmap_hum_tri}a and Fig.\ref{Fig5_2Dmap_hum_tri}b show a
qualitative similarity between real sequences of the human chromosome X and
those generated by the coupled maps (\ref{2Dcoupled}). Even in the absence of
coupling between the maps a trajectory of 2D walk has a complicated form. In
this case the dynamics of the system is only a direct product of the dynamics
of two one-dimensional maps. The 2D DNA walk of the Trypanosoma brucei chromosome 9 and  the 2D walk of the sequence generated by the model (\ref{2Dcoupled}) without coupling are shown in Fig.\ref{Fig5_2Dmap_hum_tri}c and Fig.\ref{Fig5_2Dmap_hum_tri}d, respectively. Obviously,
the presence of the diffusion coupling between the two mentioned processes may
increase the complexity of the global landscape of this sequence.

In conclusion, the results reported in this Letter show that deterministic
dynamics with intermittency can reproduce complex properties of real eukaryotic
chromosome DNA sequences including the global inverted symmetry of subtelomeres
and various types of large-scale polymorphisms. The gene number balance on DNA
strands for the most eukaryotic complete chromosome sequences (see
\cite{Poptsova2009}), which in this model are the result of the time balance of
the phase trajectory in symmetrical traps, may naturally coexist with
large-scale polymorphisms. Based on the proposed model we suppose that an increase of sequence complexity via the value of the diffusion coupling $d$
between subsystems (\ref{2Dcoupled}), represented in the dynamics of type-I and type-II intermittency, may reflect the evolutionary distance between certain system properties of real genomes.

%\bibliography{BibDNA_PRL}

\end{document}